\documentclass[10pt,twocolumn,letterpaper]{article}

\usepackage[pagenumbers]{cvpr} 

\usepackage[pagebackref,breaklinks,colorlinks]{hyperref}
\definecolor{cvprblue}{rgb}{0.21,0.49,0.74}
\definecolor{beaublue}{rgb}{0.9, 0.95, 0.9}
\definecolor{blackish}{rgb}{0.2, 0.2, 0.2}
\definecolor{greenref}{rgb}{0,0.9,0} %
\definecolor{redref}{rgb}{1,0,0} %
\definecolor{pinkref}{rgb}{0.9, 0.08, 0.58} 

\hypersetup{
    citecolor=greenref,   %
    linkcolor=redref,     %
    urlcolor=pinkref    %
}
\usepackage{tcolorbox}
\usepackage{lipsum}
\usepackage{tabularx} %
\usepackage{multirow}

\def\paperID{*****} 
\def\confName{CVPR}
\def\confYear{2025}

\title{Enhancing Monocular 3D Scene Completion with Diffusion Model}

\author{Changlin Song\thanks{Equal Contribution.} \quad Jiaqi Wang\footnotemark[1] \quad Liyun Zhu\footnotemark[1] \quad He Weng\footnotemark[1]\\
Australian National University \\
 {\tt\small  \{u7707452, u7689540, u7778917, u7705421\}@anu.edu.au} \\
}

\begin{document}
\maketitle
\begin{abstract}
3D scene reconstruction is essential for applications in virtual reality, robotics, and autonomous driving, enabling machines to understand and interact with complex environments. Traditional 3D Gaussian Splatting techniques rely on images captured from multiple viewpoints to achieve optimal performance, but this dependence limits their use in scenarios where only a single image is available. In this work, we introduce \textbf{FlashDreamer}, a novel approach for reconstructing a complete 3D scene from a single image, significantly reducing the need for multi-view inputs. Our approach leverages a pre-trained vision-language model to generate descriptive prompts for the scene, guiding a diffusion model to produce images from various perspectives, which are then fused to form a cohesive 3D reconstruction. Extensive experiments show that our method effectively and robustly expands single-image inputs into a comprehensive 3D scene, extending monocular 3D reconstruction capabilities without further training.\footnotemark Our code is available \href{https://github.com/CharlieSong1999/FlashDreamer/tree/main}{here}. 

\footnotetext{Technical report, submitted for ANU COMP 8536 project.}


\end{abstract}   

\section{Introduction}
\label{sec:intro}

3D scene reconstruction generates a three-dimensional representation of a scene from multiple input images. This essential task in computer vision provides spatial representation for applications such as autonomous driving \cite{wu2023mars}, robotics \cite{shen2023distilled}, game development \cite{game}, and virtual/augmented reality (VR/AR) \cite{zhou2018stereo}. Recently, 3D Gaussian splatting (3DGS) has gained popularity as a method for 3D representation, delivering high-quality, real-time results with minimal input \cite{charatan2024pixelsplat,10521791,kerbl20233d,li2024scenedreamer360,ma2024fastscene,rockwell2021pixelsynth}. However, 3DGS generally relies on multiple images from diverse viewpoints for optimal performance \cite{charatan2024pixelsplat,wewer2024latentsplat,yang2024gaussianobject}, limiting its adaptability for single-image scenarios \cite{MatsukiCVPR2024}. Flash3D \cite{szymanowicz2024flash3d} addresses this by enabling 3D reconstruction from a single image. However, when viewing such reconstructions from alternative angles, artifacts often appear due to insufficient information in the original image. For example, rotating the viewpoint often reveals blank areas or artifacts along the borders, as these regions lie outside the initial input.

To address these limitations, new viewpoints can be synthesized using generative models like diffusion models \cite{rombach2022high}. However, diffusion models often face consistency issues when generating multiple images of the same scene \cite{bar2024lumiere}. For instance, the overlapping regions between generated images may vary, introducing inconsistencies. Therefore, we propose \textbf{FlashDreamer}: a novel method that completes the 3D Gaussian splatting of the scene, initialized with Flash3D, by generating views from predefined angles. FlashDreamer addresses consistency by aligning overlapping areas in 3D space using intermediate 3DGS representations. Additionally, a vision-language model (VLM) provides supplementary guidance to enhance the diffusion process.
The key contributions of this work include:

\begin{itemize}
\item \textbf{FlashDreamer Pipeline:} We introduce FlashDreamer, a pipeline that constructs a complete 3D scene from an input image and surrounding viewpoints using pre-trained Flash3D, diffusion models, and VLMs.
\item \textbf{Alternative Configurations:} We explore variations within FlashDreamer, including different diffusion models, rotation angles, and prompt selections for the diffusion model.
\item \textbf{Experimental Validation:} Qualitative and quantitative evaluations demonstrate that FlashDreamer achieves effective 3D scene completion without additional training.
\end{itemize}

\begin{figure}
    \centering
    \includegraphics[width=\linewidth, trim={7.7 cm} {6.5 cm} {8.7 cm} {6.4 cm}, clip]{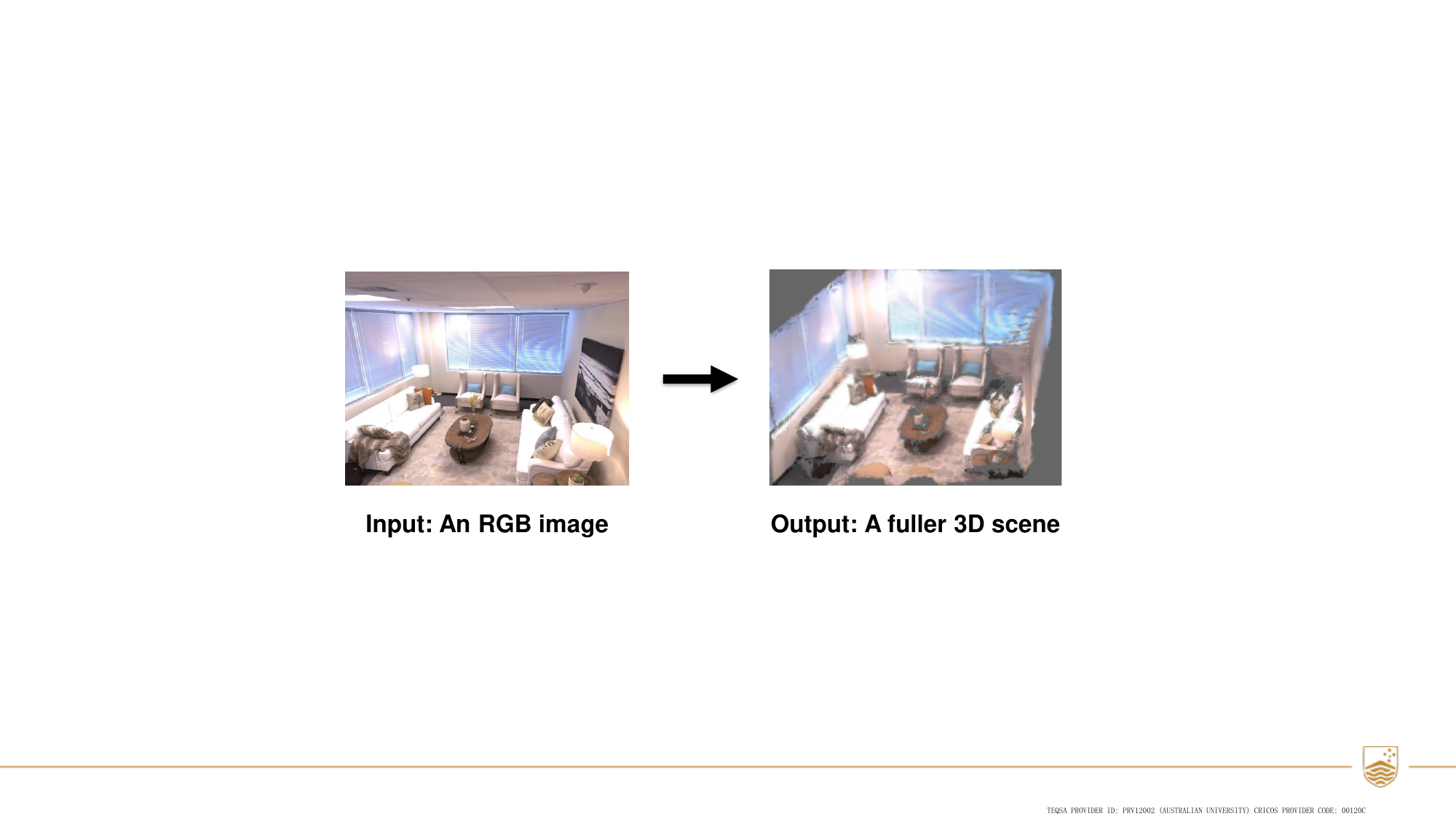}
    \caption{Motivation of our \textbf{FlashDreamer}: Given a single input image, our method reconstructs a more complete 3D scene without requiring additional images from multiple viewpoints. }
    \label{fig:motivation}
\end{figure}

\begin{figure*}[!t]
  \centering
   \includegraphics[width=1.0\linewidth]{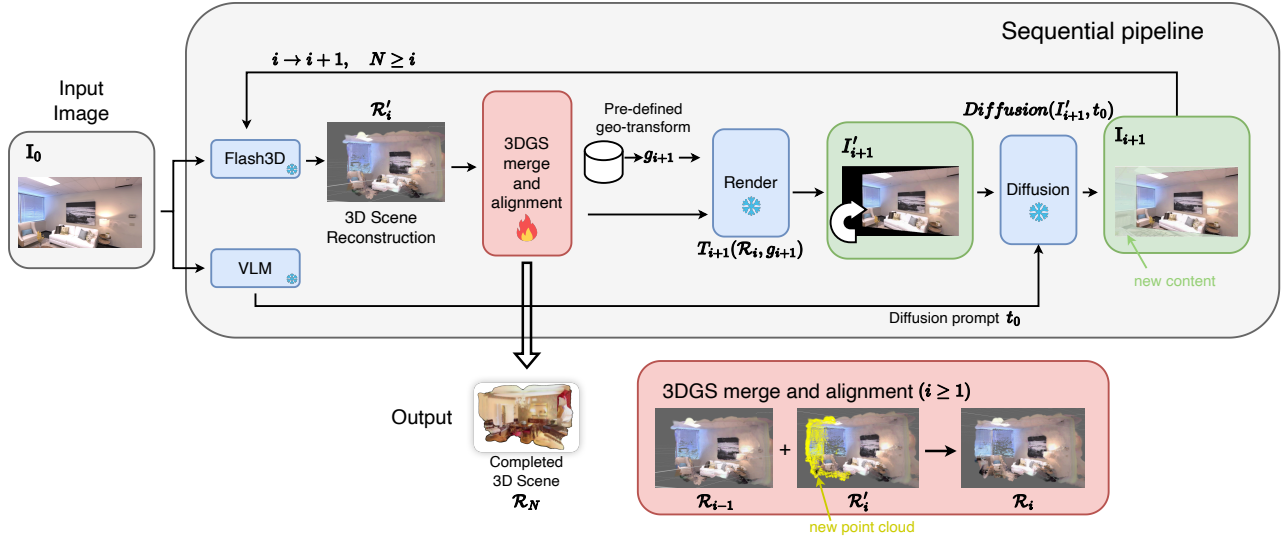}

   \caption{The pipeline of our FlashDreamer. Our model receives the initial image ($I_0$) as the input. A text prompt ($t_0$) for the diffusion inference will be generated by a pre-trained Vision-Language Model (VLM) which sees the $I_0$. The initial 3DGS ($\mathcal{R}_0$) will be generated by Flash3D, and rendered with the first pre-defined gemeotry transform $T_1$ and viewing angle $g_1$ to get the incomplete image ($I'_1 = T_1(\mathcal{R}_0, g_1)$). The diffusion model will complete $I'_1$ with inpainting, combined with the prompt $t_0$, to get $I_1$. With $I_1$ ($I_1 = Diffusion(I'_1, t_0)$), a new 3DGS $\mathcal{R}'_1$ will be generated by Flash3D, and merged with $\mathcal{R}_0$ to get the completed 3DGS ($\mathcal{R}_1$). This sequential loop will iterate $N$ times for all the pre-defined geometric transform $\{T_i\}_{i=1}^{N}$. The final 3DGS ($\mathcal{R}_N$) will be the output. }
   \label{fig:pipeline}
\end{figure*}

\section{Related Work}
\label{sec:formatting}

\noindent{\textbf{Scene reconstruction.}} Scene reconstruction refers to the process of creating a 3D representation of a real-world environment from one or more images. There have been some great effort on multiple view (image) reconstruction, from multi-view stereo \cite{schoenberger2016mvs}, Neural Radiance Field (NeRF) \cite{mildenhall2021nerf}, to 3D Gaussian Splatting (3D-GS) \cite{kerbl20233d}. Other approaches focus on reconstruction from few images \cite{charatan2024pixelsplat,wewer2024latentsplat}, demonstrating the capability to construct scenes from few viewpoints. These methods has provides accurate and real-time rendering performance in the multiple-view setting, however, these methods either require a large amount of input images or require special relationship among the input images \cite{yang2024gaussianobject}, limiting their performance in few-view or single view reconstruction. To overcome this limitation, recent study proposed two-view \cite{wang2024dust3r} and single-view reconstructors \cite{fan2024large} who leverage the transformer to capture the cross-view relationship during the training time. Flash3D \cite{szymanowicz2024flash3d} has advanced the field by offering a simple and efficient method for monocular scene reconstruction using a single image with lightweight convolution layers. 

Nonetheless, these few-view approaches can only reconstruct the content within input images, making the completion of missing content in the scene a challenge. Some methods \cite{li2024scenedreamer360,ma2024fastscene} utilize panoramic images for reconstruction, offering good loop closure due to their 360-degree coverage there no missing content within the scene. Iterative scene generation approaches, such as PixelSynth \cite{rockwell2021pixelsynth}, and Text2Room \cite{hollein2023text2room} and LucidDreamer \cite{chung2023luciddreamer}, offer another perspective by autoregressively generating scenes from partial views, or text description. Our proposed method leverages high-fidelity scene completion across multiple viewpoints while requiring only a single image as input and simple yet efficient pipeline with pre-trained models.\\

\begin{figure*}
    \centering
    \includegraphics[width=\linewidth, trim={1.6 cm} {8 cm} {1.8 cm} {8 cm}, clip]{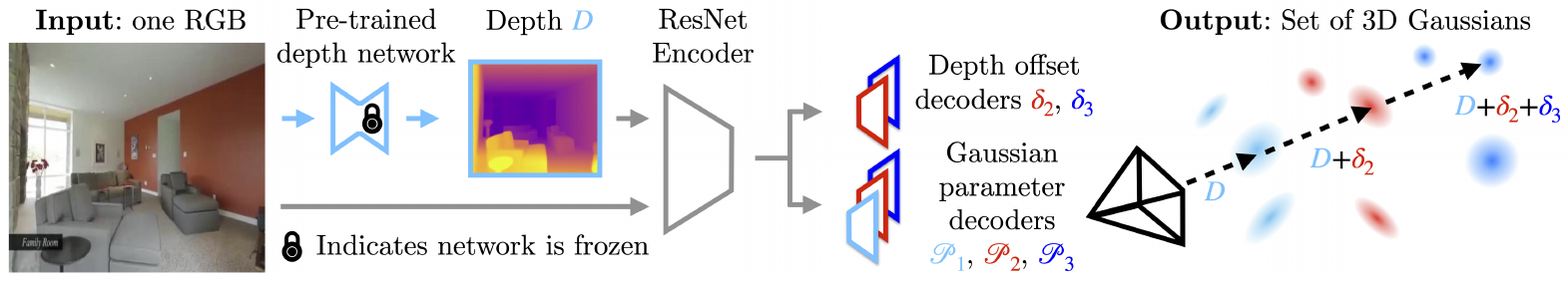}
    \caption{The pipeline of Flash3D \cite{szymanowicz2024flash3d}. Flash3D uses a ResNet encoder that extracts features from both the RGB image and its depth map estimated with a pre-trained monocular depth estimation model. They are subsequently processed by two decoders, which together output all 3D Gaussian parameters eventually. Image reproduced from \cite{szymanowicz2024flash3d}.}
    \label{fig:flash3d}
\end{figure*}

\noindent{\textbf{Diffusion model.}} The diffusion model \cite{croitoru2023diffusion} in computer vision is devised to learn the denoising process from images perturbed by Gaussian noise and then generate high-quality and diverse images. As one of the deep generative methods, diffusion-based models are widely adopted in many real-life applications such as image synthesis \cite{rombach2022high}, video generation \cite{ho2022video}, and dense visual prediction \cite{ji2023ddp}. Stable Diffusion \cite{rombach2022high,podell2023sdxl} as Latent Diffusion Models (LDMs), shows great success on the high-resolution image synthesis task. Apart from image generation, diffusion models have also gained great success in image inpainting, which is an image editing technique that can edit or complete a given image by user requirements that has the form of text description, mask, or layout \cite{huang2024diffusion}. Ledits++ \cite{brack2024ledits++} allows users to provide a mask that indicates the target region to edit, combined with certain text that describes the desired content to generate of that region. This editing pipeline aligned with our goal of completing a specific image region.  In our work, we use a pre-trained Stable Diffusion model to generate multi-view images guided by mask and text description.\\

\noindent{\textbf{Vision language model.}}
Vision Language Model (VLM) captures correlations between images and text, enabling application to downstream vision-language tasks without fine-tuning \cite{zhang2024vision}. CLIP \cite{radford2021learning}, a pioneering VLM, employs contrastive learning between text and image embeddings, each extracted by respective encoders. Its objective is to maximize similarity within matched text-image pairs while minimizing similarity across non-matching pairs. Despite CLIP’s success in matching text-image pairs, it does not address image-to-text generation. CoCa \cite{yu2022coca} addresses this by proposing an image-to-caption framework that decouples the encoder and decoder, facilitating text generation from images. In CoCa, the image and text are initially encoded separately, then cross-attended in a multimodal decoder to generate captions. In our work, we utilize LLama3 \cite{dubey2024llama}, a state-of-the-art open-source VLM, to generate textual descriptions of our pipeline based on input images.

\section{Methodology}
\label{sec:method}

\begin{figure*}
    \centering
    \includegraphics[width=\linewidth, trim={1 cm} {4.2 cm} {1 cm} {3.2 cm}, clip]{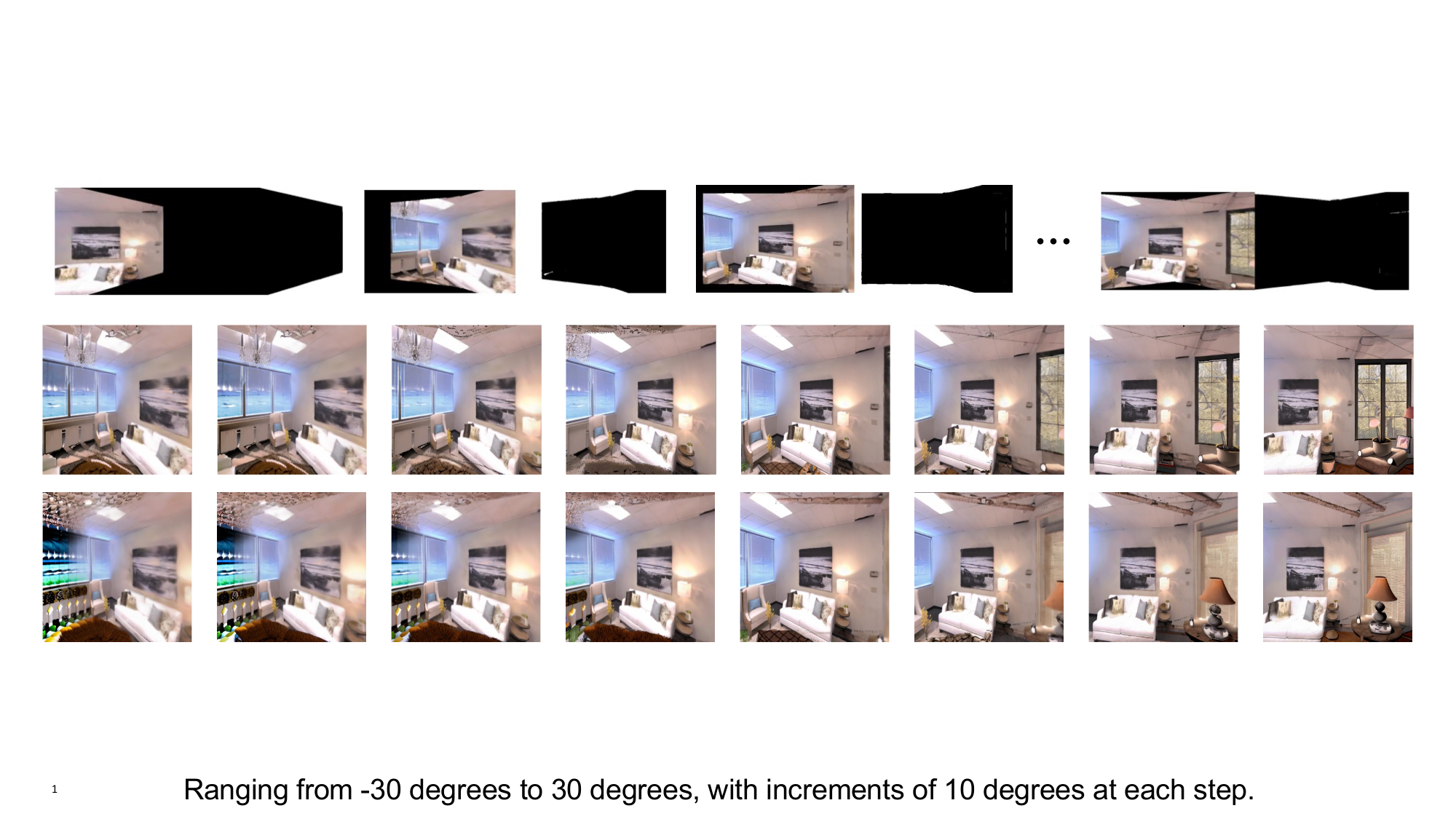}
    \caption{Comparison of different rotation angle increments. The top row presents original images rendered from various perspectives. The middle row demonstrates image rotations in 10° increments, spanning from -30° to 30°. The bottom row further refines this with 5° rotation increments over the same range. While smaller rotation increments provide finer adjustments, they result in overlapping edges, which may degrade inpainting quality by introducing artifacts in the boundary regions.
    }
    \label{fig:rotation}
\end{figure*}

\noindent{\textbf{Problem formulation.}} Our objective is to develop a pipeline capable of reconstructing a complete 3D scene using only a single image without relying on a set of images. Specifically, given a single image $I_0 \in \mathbb{R}^{H \times W \times 3}$, our method is able to generate a more complete 3D scene representation $\mathcal{R_{\text{\textit{N}}}}$. The whole pipeline is shown in \cref{fig:motivation}.
We define a collection of 3D scenes, $\{\mathcal{R}'_{i}\}_{i=0}^{N}$, generated using Flash3D~\cite{szymanowicz2024flash3d} from a single viewpoint at every step, and $\{\mathcal{R}_{i}\}_{i=0}^{N}$ denotes the accumulated 3D scene from multiple viewpoints obtained after merge and alignment at every step. Additionally, we define a set of rendering transformations $\{T_i\}_{i=0}^{N}$ and geometric transform angles $\{g_i\}_{i=0}^{N}$, to render scenes from multiple viewpoints.  Additionally, the $I'_{i} \in \mathbb{R}^{H \times W \times 3}$ denotes the image prior to inpainting while $I_{i} \in \mathbb{R}^{H \times W \times 3}$ represents the image after inpainting via a diffusion model.\\




\noindent{\textbf{FlashDreamer.}} As shown in \cref{fig:pipeline}, our pipeline involves two main steps: (i) inpainting images for new viewpoints and (ii) the 3D Gaussian Splatting (3DGS) merge and alignment process. Firstly, the input image $I_{0}$ will first be fed into the Flash3D \cite{szymanowicz2024flash3d} backbone to generate a original 3D representation of the scene $\mathcal{R}'_0$ by following equation:

\begin{equation}
\mathcal{R}'_{i} = Flash3D(I_{i})
\label{eq:1}
\end{equation}

\noindent This scene representation, $\mathcal{R}'_0$, may contain occlusions or incomplete regions due to unobserved parts of the scene. Then we render the new viewpoint image $I'_{1}$ from the original reconstructed 3D scene $\mathcal{R}_0$ \footnote{Note that for $i=0$, $\mathcal{R}_0$ = $\mathcal{R}'_0$. For $i \neq 0$, we use $\mathcal{R}_i$ for rendering.} with a pre-defined geometric transformation $g_{1}$, which means the specific angle we rotate at the first step, following:
\begin{equation}
I'_{i+1} = T_{i+1}(\mathcal{R}_{i}, g_{i+1})
\end{equation}

\noindent The black-masked
area in $I'_{1}$ indicates regions requiring inpainting by the diffusion model. Next, $I'_{1}$ is extended by the Stable Diffusion model \cite{rombach2022high} to create a extended new viewpoint image $I_{1}$. This refinement is guided by a text description $t_{0}$ of the original image $I_{0}$, which is generated by a VLM, specifically LLaMA-3.1-8B \cite{dubey2024llama}, note that to keep the consistency of each generated viewpoint, we always use the description $t_{0}$ as diffusion prompt at every step.

\begin{equation}
I_{i+1} = Diffusion(I'_{i+1}, t_0)
\end{equation}


\noindent When we obtain a inpainted scene image $I_{1}$, we reapply Eq.~\ref{eq:1} for rendering $\mathcal{R}'_1$, which is reconstructed based on the viewpoint from geometric transform angle $g_{1}$. $\mathcal{R}'_1$ will be merged and aligned with $\mathcal{R}'_0$ to get a more complete 3D representation $\mathcal{R}_1$ (see the red box in Fig.~\ref{fig:pipeline}), the merging details of which will be introduced later.


\begin{equation}
\mathcal{R}_{i+1} = Merge\&Align(\mathcal{R}_{i}, \mathcal{R}'_{i+1})
\end{equation}


\noindent The sequential pipeline goes on by applying the pre-defined geometric transform $\{g_i\}_{i=0}^{N}$ and generates the set of $\{\mathcal{R}_i\}_{i=0}^{N}$. Finally, $\mathcal{R}_N$ will be the output of the pipeline, which is the 3D representation of the scene contains both the original input image $I_0$ and images of new viewpoints $\{I_i\}_{i=1}^{N}$ generated by the diffusion model.\\

\noindent{\textbf{Flash3D.}} The scene completion process is incremental, where unobserved scenes are gradually filled with 3D Gaussians generated from Flash3D \citep{szymanowicz2024flash3d} model. The 3D coordinates of these Gaussians are first rescaled according to the dataset camera intrinsics to ensure consistency in scale. After rescaling, the coordinates are transformed from camera coordinate system into the world coordinate system. The Flash3D diagram is shown in \cref{fig:flash3d}, and the corresponding equation is simplified as \cref{eq:1}. \\

\noindent{\textbf{3D Gaussian splatting: Merge and align.}}
\noindent Given a pre-defined camera pose, we can render 3D Gaussains from new viewpoints. The newly rendered image possibly contains two parts: seen and unseen regions. Since we only need to complete unobserved part in the rendered image, a mask is created based on the existing 3D Gaussians, covering the observed regions. And the mask is defined as:
\begin{equation}
    M(u, v) = 
    \begin{cases} 
      1 & \text{if } (u, v) \text{ is observed} \\ 
      0 & \text{otherwise}
    \end{cases}
    \label{eq:mask}
\end{equation}

\noindent where $(u, v)$ are the pixel coordinates of the rendered image.\\

\noindent The mask $M$ is used as a filter for the diffusion model to complete the image based on existing RGB information. After a new image is generated, it is fed into Flash3D again, and we still use the same mask to retain the output 3D Gaussains
corresponding to pixels in the previously unobserved regions:
\begin{equation}
    \tilde{\mathcal{R}}'_{i} = \{\theta_j \in \mathcal{R}'_{i} \mid M(u_j, v_j) = 0\}
    \label{eq:filtered_gaussians}
\end{equation}

\noindent where $\mathcal{R}'_{i}$ represents the set of 3D Gaussians generated by Flash3D from image $I_i$, $\tilde {\mathcal{R}}'_{i}$ represents the 3D Gaussians we need to retain, and $(u_j, v_j)$ represents the pixel coordinates of the projection of the previous existing 3D Gaussians $\theta_j$.\\

\noindent We combine the previous and new 3D Gaussians together as an entire scene reconstruction ($\mathcal{R}_i = \mathcal{R}_{i-1} + \tilde {\mathcal{R}}'_{i}$), and optimize the 3D Gaussian parameters to make the rendered image more similar to the Flash3D inputs. The rendered images are compared with the Flash3D input images using RGB loss, defined as the pixel-wise absolute difference between the rendered image $\hat{I}_j$ and the Flash3D input image $I_j$:
\begin{equation}
    \mathcal{L}_j(\theta) = \frac{1}{HW} \sum_{u=1}^{W} \sum_{v=1}^{H} \left| \hat{I}_j(u, v) - I_j(u, v) \right|,
    \label{eq:pixelwise_loss}
\end{equation}

\noindent where $(u, v)$ represent the pixel coordinates, and the sum is taken over all pixels in the image. The overall loss function, $\mathcal{L}(\theta)$, is the average loss over all frames:
\begin{equation}
    \mathcal{L}(\theta) = \frac{1}{N} \sum_{j=1}^{N} \mathcal{L}_j(\theta),
    \label{eq:overall_loss}
\end{equation}

\noindent Eventually, we obtain a more comprehensive 3D scene reconstruction as a result of our pipeline. And we can further render it from desired viewpoints to evaluate the completion quality. \\



\section{Experiment}


In this section, we perform both qualitative and quantitative analyses to examine factors influencing scene generation quality in our method. The qualitative analysis considers (i) rotation angles, (ii) diffusion models, and (iii) prompt diversity. Quantitatively, we assess quality using Fr\'{e}chet Inception Distance (FID) \cite{heusel2017gans} and CLIP Score \cite{radford2021learning} across rotation angles, aiming to clarify key parameters that drive high-quality scene generation.\\


\noindent{\textbf{Setup.}} We employ Flash3D \cite{szymanowicz2024flash3d} for efficient scene reconstruction, combined with a pre-trained Stable Diffusion-v2 model \cite{Rombach_2022_CVPR} to expand scene images across multiple viewpoints. To enhance the generation accuracy of the diffusion model, we input a guiding prompt, "Please briefly describe the scene", into the LLaMA-3.1-8B \cite{dubey2024llama}. The visual language model (VLM) then generates a description, which serves as the prompt for the diffusion model. Finally, a standard 3DGS pipeline is applied to reconstruct the 3D scene. Due to the time limits, our experiments are conducted on a subset of the Replica dataset \cite{straub2019replica}, which contains image frames for 18 highly photorealistic 3D indoor scenes, providing a diverse and comprehensive simulation of real-world indoor scenes. The subset contains 20 images randomly selected from the replica dataset, and for each image, we chose 6 new viewing angles ranging from -30° to 30° with 10° as rotation unit, \textit{i.e.}, the subset contains 20 input images, and 120 ground-truth images. 
To ensure efficient processing and optimal performance, we utilize an NVIDIA Tesla V100 GPU for all experiments.\\

\noindent{\textbf{Baseline.}} PixelSynth \cite{rockwell2021pixelsynth} is a monocular scene completion models that use a generative model to complete the content beyond the input image, and serves as the baseline to compare with our model. The difference is that they did not utilize a 3D representation of the scene, and trained a GAN \cite{heusel2017gans} to generate the pixel content. In our model, we use Flash3D to generate intermediate 3DGS for 3D consistency maintenance, and we use a pre-trained diffusion model and VLMs that are more generalized than a trained GAN.\\

\noindent{\textbf{Evaluation metrics.}} Considering the generative nature of the task, we adopt two metrics for generative task including: (i) Fr\'{e}chet Inception Distance(FID) \cite{heusel2017gans}, measuring the distribution drift between generated and real data. (ii) CLIP Score \cite{radford2021learning}, measuring the similarity of generative image with text prompt. For quantitative evaluation regarding the metrics of CLIP score and FID, we use the API \footnote{\href{https://github.com/facebookresearch/Replica-Dataset}{https://github.com/facebookresearch/Replica-Dataset}} to obtain images of 3D-rendering scenes as the inputs, along with images from various camera rotation angles as the ground truth. Then, we feed inputs into methods and request them to complete the missing parts of the original image rotated by preset angles.\\


\begin{figure}
    \centering
    \includegraphics[width=\linewidth, trim={8.0 cm} {4.7 cm} {7.0 cm} {4.2 cm}, clip]{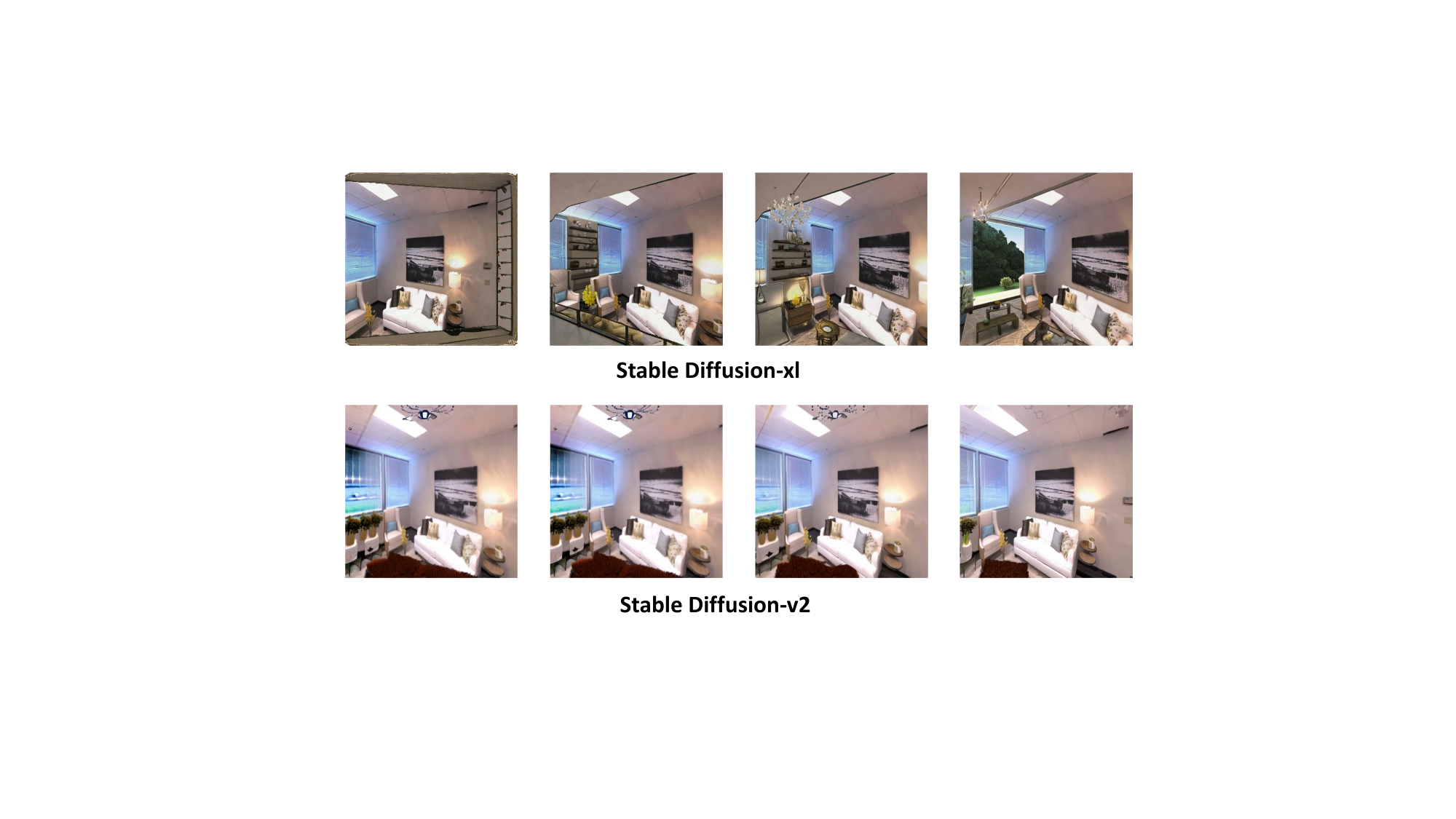}
    \caption{Comparison of diffusion models. Images generated with identical prompts and using Stable Diffusion-v2~\cite{Rombach_2022_CVPR} and Stable Diffusion-xl~\cite{podell2023sdxl}. Stable Diffusion-v2 yields more realistic outputs, while Stable Diffusion-xl shows inconsistencies and artifacts.}
    \label{fig:diffusion}
\end{figure}

\begin{figure}
    \centering
    \includegraphics[width=\linewidth, trim={8 cm} {4.2 cm} {7 cm} {4.9 cm}, clip]{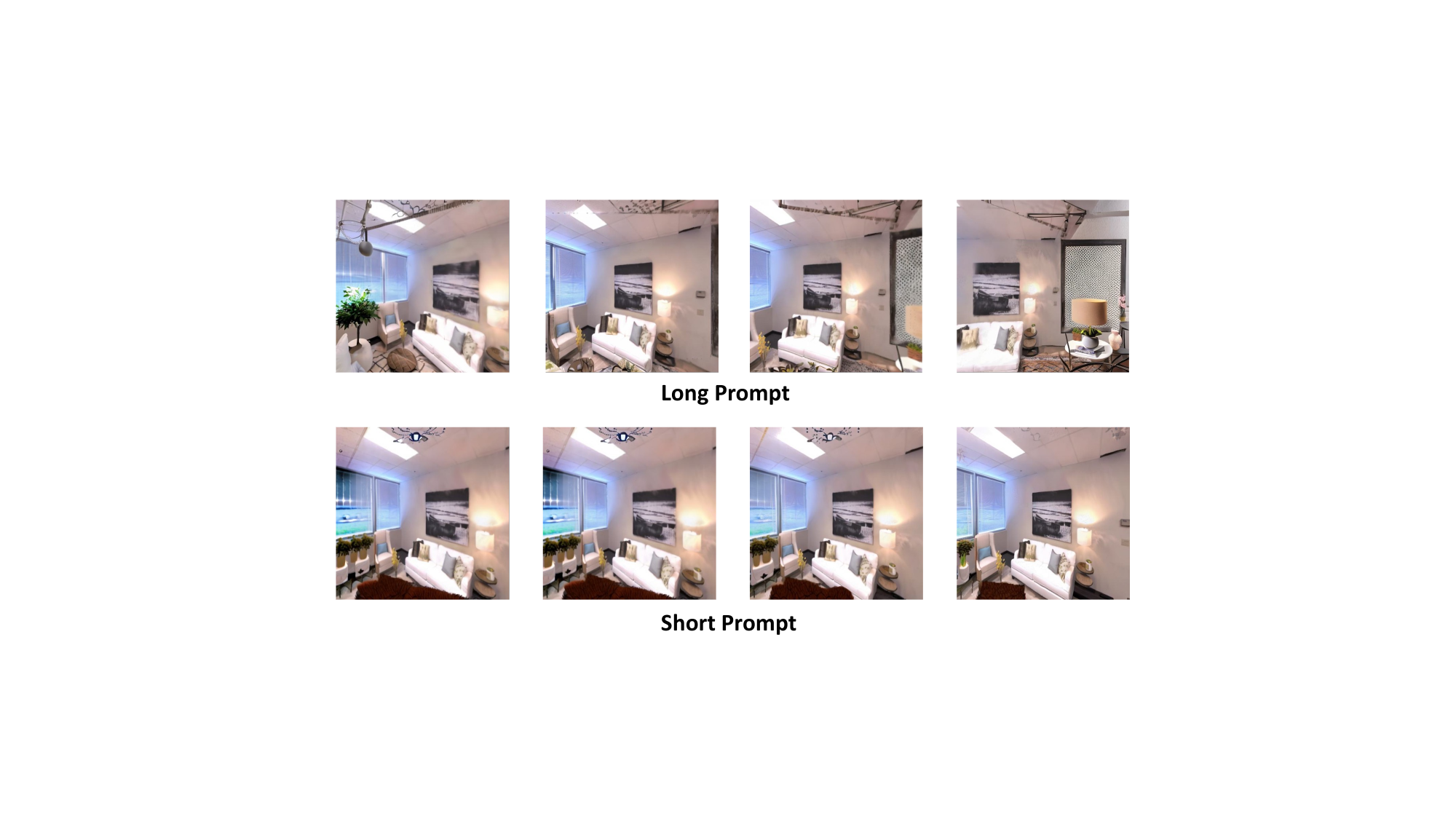}
    \caption{Comparison between diffusion prompts. Longer prompts enhance detail but introduce more artifacts and reducing structural consistency. Shorter prompts create simpler layouts but maintain stronger structural consistency.}
    \label{fig:prompt}
\end{figure}

\noindent{\textbf{Rotation angle.}} 
As illustrated in the first row of Fig.~\ref{fig:rotation}, our experiment demonstrates the process of rendering original images from multiple perspectives, with black-masked areas indicating regions that the diffusion model needs to inpaint. By combining inpainted images from different angles, we construct the final 3D scene, making the selection of rotation angle increments crucial to the model's performance. Our study explores the impact of two rotation increments, 5° and 10°, on diffusion model generation quality, with angles ranging from -30° to 30°, as shown in the middle and bottom rows of Fig.~\ref{fig:rotation}. The results reveal that smaller increments, while providing finer adjustments, lead to edge overlap across viewpoints, which introduces more artifacts that can hinder inpainting quality.\\

\noindent{\textbf{Comparison between diffusion models.}}  Diffusion models can exhibit diverse visual characteristics even when using identical rotation angles and prompts, highlighting their inherent stylistic differences. In our experiment, we utilize Stable Diffusion-v2~\cite{Rombach_2022_CVPR} and Stable Diffusion-xl~\cite{podell2023sdxl}. As demonstrated in Fig.~\ref{fig:diffusion}, Stable Diffusion-v2 achieves more photographic results compared to Stable Diffusion-xl. The quality can be advantageous for applications that prioritize realism across multiple views. Conversely, Stable Diffusion-xl, while more flexible in producing diverse and artistic visual styles, may introduce elements that deviate stylistically from the source images. The added artistic capabilities of Stable Diffusion-xl can be beneficial for creative applications but may result in inconsistencies in style when compared to the original images, particularly in scenarios requiring uniformity.\\


\noindent{\textbf{Comparison between prompts.}}
Existing powerful Vision-Language Models (VLMs) can elegantly describe the features of a scene~\cite{zhang2024vision}. However, when we input these prompts into the diffusion model, we encounter two issues: (i) the token length that can be input into the diffusion model is limited, and (ii) a well-crafted descriptive prompt does not necessarily yield high-quality scene generation.
Therefore, we conduct experiments on Stable Diffusion-v2~\cite{Rombach_2022_CVPR} to identify the most effective prompt types for scene generation. To investigate how prompt detail affects the generated results, we use two types of diffusion prompts, labeled in the green box as the ``short prompt" and the ``long prompt". The long prompt describes specific items in the room, their relative positions, and additional scene elements, providing the model with a richer context. While the short prompt shortly describes the items in the scene. 

\begin{tcolorbox}[width=1.0\linewidth, colframe=blackish, colback=beaublue, boxsep=0mm, arc=3mm, left=1mm, right=1mm, right=1mm, top=1mm, bottom=1mm]
\textbf{Short prompt:} An indoor scene, a window, two sofas.

\textbf{Long prompt:} The room features a white sofa with several pillows. To the left of the sofa, there is an armchair with a blue cushion, and in front of the sofa is a small round wooden table with a decorative plant in a vase. On the right side, there is a round side table with two wooden tiers, topped with a small lamp. The lamp casts a warm glow on the wall. Above the sofa, a large black-and-white framed photograph of a canoe by a lake or river hangs on the wall, adding a nature-inspired element to the space. The walls are painted a light neutral color, and the room has a drop ceiling with overhead lighting.
\end{tcolorbox}


\noindent As shown in Fig.~\ref{fig:prompt}, images generated with longer prompts exhibit enhanced detail, capturing additional elements such as lamps and decorative plants, which improves texture and object fidelity. However, these images tend to lack structural consistency and contain more artifacts. In contrast, images generated with shorter prompts present a simpler layout, with fewer scene details but greater structural consistency. This observation highlights the role of prompt engineering in single-image 3D scene reconstruction, where a carefully tailored prompt can effectively guide the model to produce a more complete and realistic scene.\\

\begin{figure}
    \centering
    \includegraphics[width=\linewidth, trim={9.7 cm} {4.7 cm} {10.3 cm} {3.5 cm}, clip]{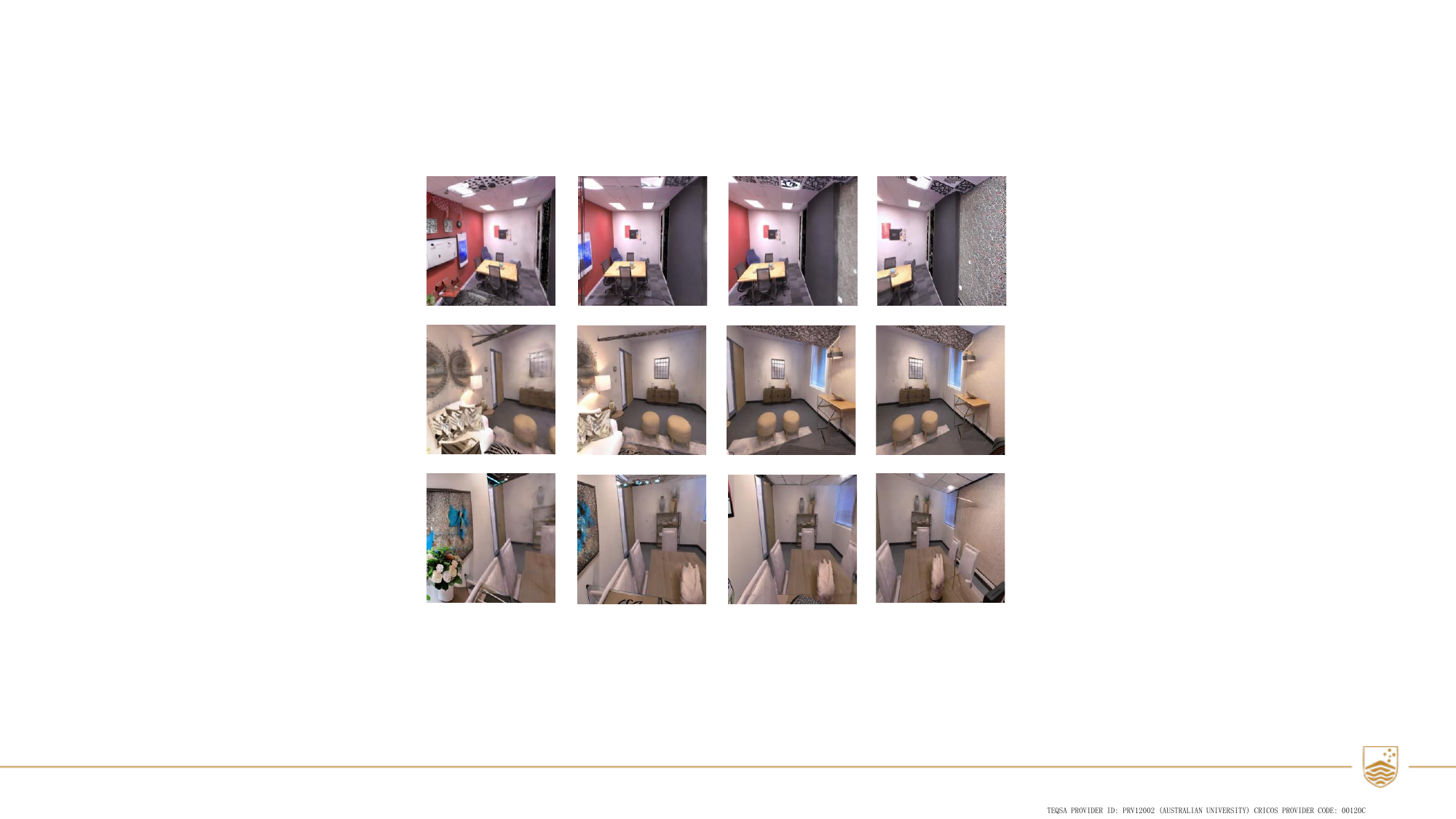}
    \caption{Additional inpainting results generated using our method with Stable Diffusion-v2. The images demonstrate the model's capability to capture spatial coherence across multiple scenes.}
    \label{fig:visualization}
\end{figure}

\begin{table}[tbp]

\caption{Comparison of Fr\'{e}chet Inception Distance(FID) \cite{heusel2017gans} and CLIP Score \cite{radford2021learning} for Pixelsynth (PixSyn)~\cite{rockwell2021pixelsynth} and our FlashDreamer (FD) across various horizontal  rotation angles. Lower FID and higher CLIP scores indicate better performance.}
\resizebox{\columnwidth}{!}{
\begin{tabular}{ccccccc}
\toprule

\multirow{2.5}{*}{Method} & \multicolumn{2}{c}{-30°} & \multicolumn{2}{c}{-20°} & \multicolumn{2}{c}{-10°} \\ 
\cmidrule{2-3}  \cmidrule{4-5} \cmidrule{6-7}
&FID $\downarrow$ & CLIP $\uparrow$ &FID $\downarrow$ & CLIP $\uparrow$&FID $\downarrow$ & CLIP $\uparrow$ \\ 
\midrule
PixSyn~\cite{rockwell2021pixelsynth} & 252.54 & 0.251 & \textbf{229.37} & 0.261 & 207.26 & 0.260  \\
\textbf{FD (ours)} & \textbf{247.31} & \textbf{0.266} & 232.46 & \textbf{0.273} & \textbf{175.39} & \textbf{0.268}  \\ 

\midrule

\multirow{2.5}{*}{Method} & \multicolumn{2}{c}{10°} & \multicolumn{2}{c}{20°} & \multicolumn{2}{c}{30°} \\ 
\cmidrule{2-3}  \cmidrule{4-5} \cmidrule{6-7}
&FID $\downarrow$ & CLIP $\uparrow$ &FID $\downarrow$ & CLIP $\uparrow$&FID $\downarrow$ & CLIP $\uparrow$ \\ 
\midrule

PixSyn~\cite{rockwell2021pixelsynth} & 216.94 & 0.255 & \textbf{223.96} & 0.252 & \textbf{228.28} & 0.245 \\
\textbf{FD (ours)} & \textbf{186.36} & \textbf{0.262} & 227.37 & \textbf{0.273} & 243.49 & \textbf{0.266} \\
\bottomrule
\end{tabular}
}
\label{tab:result}
\end{table}

\noindent{\textbf{Quantitative results.}} As shown in Table.~\ref{tab:result}, our experiment compares the Fréchet Inception Distance (FID) and CLIP scores between two methods, PixelSynth and our FlashDreamer, at different rotation angles. FID evaluates the quality of generated images, where lower scores indicate better image fidelity, while CLIP score measures alignment with text prompts, with higher scores being preferable. As the absolute value of the rotation angle decreases, both methods demonstrate progressively lower FID scores, indicating improved image quality. Similarly, CLIP scores increase at these smaller angles, suggesting better prompt alignment. Overall, these results highlight that smaller rotation angles contribute to higher-quality image generation, with our FlashDreamer outperforming PixelSynth across most evaluated angles.\\


\begin{figure}
    \centering
    \includegraphics[width=\linewidth, trim={11.3 cm} {6.2 cm} {10.8 cm} {6.5 cm}, clip]{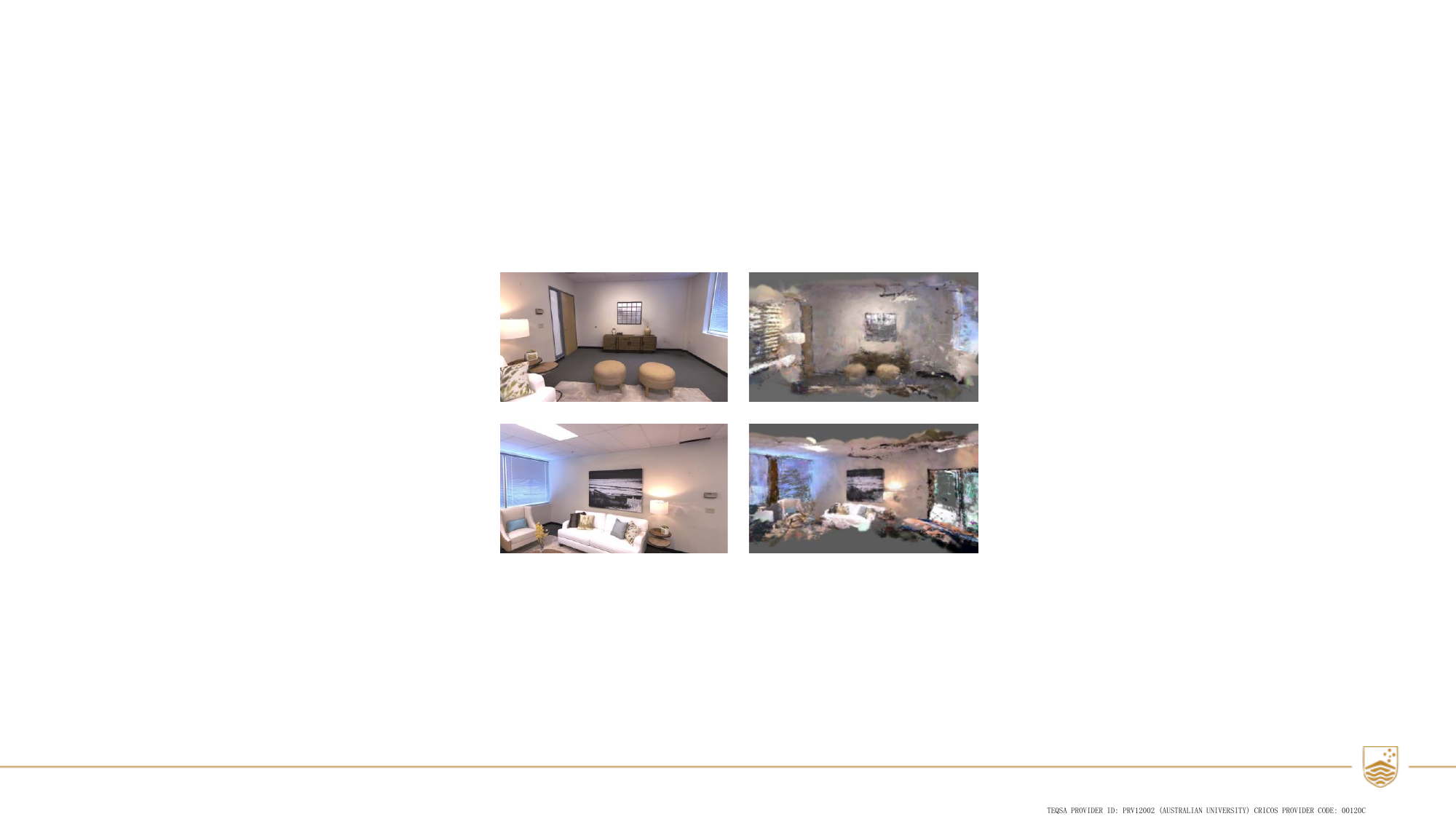}
    \caption{Scene completion results using our method. The left column shows the original input images, while the right column displays the reconstructed 3D scenes.}
    \label{fig:visualization-2}
\end{figure}

\noindent{\textbf{Further visualization.}}  We present inpainted room images from the Replica dataset in Fig.~\ref{fig:visualization}. The generated images showcase various perspectives and settings, capturing different environments such as offices, living rooms, and meeting spaces. Additionally, Fig.~\ref{fig:visualization-2} shows examples of our 3D reconstructions, demonstrating structural completeness and spatial depth that clarify room geometry. These results highlight the challenges of capturing fine details and keeping textures consistent across different views.

\section{Conclusion}


We propose FlashDreamer, a new method that advances monocular 3D scene reconstruction by creating complete 3D environments from a single image, removing the need for multi-view images required by traditional 3D Gaussian Splatting. By leveraging a vision-language model to generate descriptive prompts that guide a diffusion model in producing multi-perspective images, FlashDreamer achieves accurate and cohesive 3D reconstructions. Our approach requires no additional training, highlighting its efficiency and adaptability across applications in virtual reality, robotics, and autonomous driving. Extensive experiments confirm that FlashDreamer robustly transforms single-image inputs into comprehensive 3D scenes, advancing the field of single-image 3D reconstruction.



\section{Acknowledgment}

This work represents an equal contribution from Changlin Song, Jiaqi Wang, Liyun Zhu, and He Weng. We deeply appreciate everyone’s dedication and hard work, and it has been a truly enjoyable and rewarding collaboration. The primary contributions from each team member are as follows: Changlin Song contributed to drafting the proposal, conducting the literature review, constructing the pipeline code, and writing the report sections on introduction and methods. Jiaqi Wang implemented the Flash3D code, contributed to the literature review, and wrote the section on methods specific to Flash3D. Liyun Zhu contributed to drafting the proposal, conducted the visualization and comparison experiments, and authored sections on methods, experiments, and conclusion. He Weng focused on the literature review, ran experiments related to VLM and the diffusion model, and conducted quantitative evaluations.


{
    \small
    \bibliographystyle{ieeenat_fullname}
    \bibliography{main}

\begin{thebibliography}{35}
\providecommand{\natexlab}[1]{#1}
\providecommand{\url}[1]{\texttt{#1}}
\expandafter\ifx\csname urlstyle\endcsname\relax
  \providecommand{\doi}[1]{doi: #1}\else
  \providecommand{\doi}{doi: \begingroup \urlstyle{rm}\Url}\fi

\bibitem[Bar-Tal et~al.(2024)Bar-Tal, Chefer, Tov, Herrmann, Paiss, Zada, Ephrat, Hur, Liu, Raj, et~al.]{bar2024lumiere}
Omer Bar-Tal, Hila Chefer, Omer Tov, Charles Herrmann, Roni Paiss, Shiran Zada, Ariel Ephrat, Junhwa Hur, Guanghui Liu, Amit Raj, et~al.
\newblock Lumiere: A space-time diffusion model for video generation.
\newblock \emph{arXiv preprint arXiv:2401.12945}, 2024.

\bibitem[Brack et~al.(2024)Brack, Friedrich, Kornmeier, Tsaban, Schramowski, Kersting, and Passos]{brack2024ledits++}
Manuel Brack, Felix Friedrich, Katharia Kornmeier, Linoy Tsaban, Patrick Schramowski, Kristian Kersting, and Apolin{\'a}rio Passos.
\newblock Ledits++: Limitless image editing using text-to-image models.
\newblock In \emph{Proceedings of the IEEE/CVF Conference on Computer Vision and Pattern Recognition}, pages 8861--8870, 2024.

\bibitem[Charatan et~al.(2024)Charatan, Li, Tagliasacchi, and Sitzmann]{charatan2024pixelsplat}
David Charatan, Sizhe~Lester Li, Andrea Tagliasacchi, and Vincent Sitzmann.
\newblock pixelsplat: 3d gaussian splats from image pairs for scalable generalizable 3d reconstruction.
\newblock In \emph{Proceedings of the IEEE/CVF Conference on Computer Vision and Pattern Recognition}, pages 19457--19467, 2024.

\bibitem[Chung et~al.(2023)Chung, Lee, Nam, Lee, and Lee]{chung2023luciddreamer}
Jaeyoung Chung, Suyoung Lee, Hyeongjin Nam, Jaerin Lee, and Kyoung~Mu Lee.
\newblock Luciddreamer: Domain-free generation of 3d gaussian splatting scenes.
\newblock \emph{arXiv preprint arXiv:2311.13384}, 2023.

\bibitem[Croitoru et~al.(2023)Croitoru, Hondru, Ionescu, and Shah]{croitoru2023diffusion}
Florinel-Alin Croitoru, Vlad Hondru, Radu~Tudor Ionescu, and Mubarak Shah.
\newblock Diffusion models in vision: A survey.
\newblock \emph{IEEE Transactions on Pattern Analysis and Machine Intelligence}, 45\penalty0 (9):\penalty0 10850--10869, 2023.

\bibitem[Dubey et~al.(2024)Dubey, Jauhri, Pandey, Kadian, Al-Dahle, Letman, Mathur, Schelten, Yang, Fan, et~al.]{dubey2024llama}
Abhimanyu Dubey, Abhinav Jauhri, Abhinav Pandey, Abhishek Kadian, Ahmad Al-Dahle, Aiesha Letman, Akhil Mathur, Alan Schelten, Amy Yang, Angela Fan, et~al.
\newblock The llama 3 herd of models.
\newblock \emph{arXiv preprint arXiv:2407.21783}, 2024.

\bibitem[Fan et~al.(2024)Fan, Zhang, Cong, Wang, Li, Wen, Zhou, Kadambi, Wang, Xu, et~al.]{fan2024large}
Zhiwen Fan, Jian Zhang, Wenyan Cong, Peihao Wang, Renjie Li, Kairun Wen, Shijie Zhou, Achuta Kadambi, Zhangyang Wang, Danfei Xu, et~al.
\newblock Large spatial model: End-to-end unposed images to semantic 3d.
\newblock \emph{arXiv preprint arXiv:2410.18956}, 2024.

\bibitem[Fei et~al.(2024)Fei, Xu, Zhang, Zhou, Yang, and He]{10521791}
Ben Fei, Jingyi Xu, Rui Zhang, Qingyuan Zhou, Weidong Yang, and Ying He.
\newblock 3d gaussian splatting as new era: A survey.
\newblock \emph{IEEE Transactions on Visualization and Computer Graphics}, pages 1--20, 2024.

\bibitem[Heusel et~al.(2017)Heusel, Ramsauer, Unterthiner, Nessler, and Hochreiter]{heusel2017gans}
Martin Heusel, Hubert Ramsauer, Thomas Unterthiner, Bernhard Nessler, and Sepp Hochreiter.
\newblock Gans trained by a two time-scale update rule converge to a local nash equilibrium.
\newblock \emph{Advances in neural information processing systems}, 30, 2017.

\bibitem[Ho et~al.(2022)Ho, Salimans, Gritsenko, Chan, Norouzi, and Fleet]{ho2022video}
Jonathan Ho, Tim Salimans, Alexey Gritsenko, William Chan, Mohammad Norouzi, and David~J Fleet.
\newblock Video diffusion models.
\newblock \emph{Advances in Neural Information Processing Systems}, 35:\penalty0 8633--8646, 2022.

\bibitem[H{\"o}llein et~al.(2023)H{\"o}llein, Cao, Owens, Johnson, and Nie{\ss}ner]{hollein2023text2room}
Lukas H{\"o}llein, Ang Cao, Andrew Owens, Justin Johnson, and Matthias Nie{\ss}ner.
\newblock Text2room: Extracting textured 3d meshes from 2d text-to-image models.
\newblock In \emph{Proceedings of the IEEE/CVF International Conference on Computer Vision}, pages 7909--7920, 2023.

\bibitem[Huang et~al.(2024)Huang, Huang, Liu, Yan, Lv, Liu, Xiong, Zhang, Chen, and Cao]{huang2024diffusion}
Yi Huang, Jiancheng Huang, Yifan Liu, Mingfu Yan, Jiaxi Lv, Jianzhuang Liu, Wei Xiong, He Zhang, Shifeng Chen, and Liangliang Cao.
\newblock Diffusion model-based image editing: A survey.
\newblock \emph{arXiv preprint arXiv:2402.17525}, 2024.

\bibitem[Ji et~al.(2023)Ji, Chen, Xie, Hong, Liu, Liu, Lu, Li, and Luo]{ji2023ddp}
Yuanfeng Ji, Zhe Chen, Enze Xie, Lanqing Hong, Xihui Liu, Zhaoqiang Liu, Tong Lu, Zhenguo Li, and Ping Luo.
\newblock Ddp: Diffusion model for dense visual prediction.
\newblock In \emph{Proceedings of the IEEE/CVF International Conference on Computer Vision}, pages 21741--21752, 2023.

\bibitem[Kerbl et~al.(2023)Kerbl, Kopanas, Leimk{\"u}hler, and Drettakis]{kerbl20233d}
Bernhard Kerbl, Georgios Kopanas, Thomas Leimk{\"u}hler, and George Drettakis.
\newblock 3d gaussian splatting for real-time radiance field rendering.
\newblock \emph{ACM Trans. Graph.}, 42\penalty0 (4):\penalty0 139--1, 2023.

\bibitem[Li et~al.(2024)Li, Mi, Cai, Yang, Zuo, Wang, and Fan]{li2024scenedreamer360}
Wenrui Li, Yapeng Mi, Fucheng Cai, Zhe Yang, Wangmeng Zuo, Xingtao Wang, and Xiaopeng Fan.
\newblock Scenedreamer360: Text-driven 3d-consistent scene generation with panoramic gaussian splatting.
\newblock \emph{arXiv preprint arXiv:2408.13711}, 2024.

\bibitem[Ma et~al.(2024)Ma, Zhan, and Jin]{ma2024fastscene}
Yikun Ma, Dandan Zhan, and Zhi Jin.
\newblock Fastscene: Text-driven fast 3d indoor scene generation via panoramic gaussian splatting.
\newblock \emph{arXiv preprint arXiv:2405.05768}, 2024.

\bibitem[Matsuki et~al.(2024)Matsuki, Murai, Kelly, and Davison]{MatsukiCVPR2024}
Hidenobu Matsuki, Riku Murai, Paul H.~J. Kelly, and Andrew~J. Davison.
\newblock Gaussian splatting slam.
\newblock In \emph{Proceedings of the IEEE/CVF Conference on Computer Vision and Pattern Recognition}, 2024.

\bibitem[Mildenhall et~al.(2021)Mildenhall, Srinivasan, Tancik, Barron, Ramamoorthi, and Ng]{mildenhall2021nerf}
Ben Mildenhall, Pratul~P Srinivasan, Matthew Tancik, Jonathan~T Barron, Ravi Ramamoorthi, and Ren Ng.
\newblock Nerf: Representing scenes as neural radiance fields for view synthesis.
\newblock \emph{Communications of the ACM}, 65\penalty0 (1):\penalty0 99--106, 2021.

\bibitem[Podell et~al.(2023)Podell, English, Lacey, Blattmann, Dockhorn, M{\"u}ller, Penna, and Rombach]{podell2023sdxl}
Dustin Podell, Zion English, Kyle Lacey, Andreas Blattmann, Tim Dockhorn, Jonas M{\"u}ller, Joe Penna, and Robin Rombach.
\newblock Sdxl: Improving latent diffusion models for high-resolution image synthesis.
\newblock \emph{arXiv preprint arXiv:2307.01952}, 2023.

\bibitem[Radford et~al.(2021)Radford, Kim, Hallacy, Ramesh, Goh, Agarwal, Sastry, Askell, Mishkin, Clark, et~al.]{radford2021learning}
Alec Radford, Jong~Wook Kim, Chris Hallacy, Aditya Ramesh, Gabriel Goh, Sandhini Agarwal, Girish Sastry, Amanda Askell, Pamela Mishkin, Jack Clark, et~al.
\newblock Learning transferable visual models from natural language supervision.
\newblock In \emph{International conference on machine learning}, pages 8748--8763. PMLR, 2021.

\bibitem[Rockwell et~al.(2021)Rockwell, Fouhey, and Johnson]{rockwell2021pixelsynth}
Chris Rockwell, David~F Fouhey, and Justin Johnson.
\newblock Pixelsynth: Generating a 3d-consistent experience from a single image.
\newblock In \emph{Proceedings of the IEEE/CVF International Conference on Computer Vision}, pages 14104--14113, 2021.

\bibitem[Rombach et~al.(2022{\natexlab{a}})Rombach, Blattmann, Lorenz, Esser, and Ommer]{Rombach_2022_CVPR}
Robin Rombach, Andreas Blattmann, Dominik Lorenz, Patrick Esser, and Bj\"orn Ommer.
\newblock High-resolution image synthesis with latent diffusion models.
\newblock In \emph{Proceedings of the IEEE/CVF Conference on Computer Vision and Pattern Recognition (CVPR)}, pages 10684--10695, 2022{\natexlab{a}}.

\bibitem[Rombach et~al.(2022{\natexlab{b}})Rombach, Blattmann, Lorenz, Esser, and Ommer]{rombach2022high}
Robin Rombach, Andreas Blattmann, Dominik Lorenz, Patrick Esser, and Bj{\"o}rn Ommer.
\newblock High-resolution image synthesis with latent diffusion models.
\newblock In \emph{Proceedings of the IEEE/CVF conference on computer vision and pattern recognition}, pages 10684--10695, 2022{\natexlab{b}}.

\bibitem[Sch\"{o}nberger et~al.(2016)Sch\"{o}nberger, Zheng, Pollefeys, and Frahm]{schoenberger2016mvs}
Johannes~Lutz Sch\"{o}nberger, Enliang Zheng, Marc Pollefeys, and Jan-Michael Frahm.
\newblock Pixelwise view selection for unstructured multi-view stereo.
\newblock In \emph{European Conference on Computer Vision (ECCV)}, 2016.

\bibitem[Shen et~al.(2023)Shen, Yang, Yu, Wong, Kaelbling, and Isola]{shen2023distilled}
William Shen, Ge Yang, Alan Yu, Jansen Wong, Leslie~Pack Kaelbling, and Phillip Isola.
\newblock Distilled feature fields enable few-shot language-guided manipulation.
\newblock \emph{arXiv preprint arXiv:2308.07931}, 2023.

\bibitem[Straub et~al.(2019)Straub, Whelan, Ma, Chen, Wijmans, Green, Engel, Mur-Artal, Ren, Verma, et~al.]{straub2019replica}
Julian Straub, Thomas Whelan, Lingni Ma, Yufan Chen, Erik Wijmans, Simon Green, Jakob~J Engel, Raul Mur-Artal, Carl Ren, Shobhit Verma, et~al.
\newblock The replica dataset: A digital replica of indoor spaces.
\newblock \emph{arXiv preprint arXiv:1906.05797}, 2019.

\bibitem[Szwoch and Bartoszewski(2020)]{game}
Mariusz Szwoch and Dariusz Bartoszewski.
\newblock 3d optical reconstruction of building interiors for game development.
\newblock In \emph{Image Processing and Communications}, pages 114--124, Cham, 2020. Springer International Publishing.

\bibitem[Szymanowicz et~al.(2024)Szymanowicz, Insafutdinov, Zheng, Campbell, Henriques, Rupprecht, and Vedaldi]{szymanowicz2024flash3d}
Stanislaw Szymanowicz, Eldar Insafutdinov, Chuanxia Zheng, Dylan Campbell, Jo{\~a}o~F Henriques, Christian Rupprecht, and Andrea Vedaldi.
\newblock Flash3d: Feed-forward generalisable 3d scene reconstruction from a single image.
\newblock \emph{arXiv preprint arXiv:2406.04343}, 2024.

\bibitem[Wang et~al.(2024)Wang, Leroy, Cabon, Chidlovskii, and Revaud]{wang2024dust3r}
Shuzhe Wang, Vincent Leroy, Yohann Cabon, Boris Chidlovskii, and Jerome Revaud.
\newblock Dust3r: Geometric 3d vision made easy.
\newblock In \emph{Proceedings of the IEEE/CVF Conference on Computer Vision and Pattern Recognition}, pages 20697--20709, 2024.

\bibitem[Wewer et~al.(2024)Wewer, Raj, Ilg, Schiele, and Lenssen]{wewer2024latentsplat}
Christopher Wewer, Kevin Raj, Eddy Ilg, Bernt Schiele, and Jan~Eric Lenssen.
\newblock latentsplat: Autoencoding variational gaussians for fast generalizable 3d reconstruction.
\newblock \emph{arXiv preprint arXiv:2403.16292}, 2024.

\bibitem[Wu et~al.(2023)Wu, Liu, Luo, Zhong, Chen, Xiao, Hou, Lou, Chen, Yang, et~al.]{wu2023mars}
Zirui Wu, Tianyu Liu, Liyi Luo, Zhide Zhong, Jianteng Chen, Hongmin Xiao, Chao Hou, Haozhe Lou, Yuantao Chen, Runyi Yang, et~al.
\newblock Mars: An instance-aware, modular and realistic simulator for autonomous driving.
\newblock In \emph{CAAI International Conference on Artificial Intelligence}, pages 3--15. Springer, 2023.

\bibitem[Yang et~al.(2024)Yang, Li, Fang, Liang, Xie, Zhang, Shen, and Tian]{yang2024gaussianobject}
Chen Yang, Sikuang Li, Jiemin Fang, Ruofan Liang, Lingxi Xie, Xiaopeng Zhang, Wei Shen, and Qi Tian.
\newblock Gaussianobject: Just taking four images to get a high-quality 3d object with gaussian splatting.
\newblock \emph{arXiv preprint arXiv:2402.10259}, 2024.

\bibitem[Yu et~al.(2022)Yu, Wang, Vasudevan, Yeung, Seyedhosseini, and Wu]{yu2022coca}
Jiahui Yu, Zirui Wang, Vijay Vasudevan, Legg Yeung, Mojtaba Seyedhosseini, and Yonghui Wu.
\newblock Coca: Contrastive captioners are image-text foundation models.
\newblock \emph{Transactions on Machine Learning Research}, 2022.

\bibitem[Zhang et~al.(2024)Zhang, Huang, Jin, and Lu]{zhang2024vision}
Jingyi Zhang, Jiaxing Huang, Sheng Jin, and Shijian Lu.
\newblock Vision-language models for vision tasks: A survey.
\newblock \emph{IEEE Transactions on Pattern Analysis and Machine Intelligence}, 2024.

\bibitem[Zhou et~al.(2018)Zhou, Tucker, Flynn, Fyffe, and Snavely]{zhou2018stereo}
Tinghui Zhou, Richard Tucker, John Flynn, Graham Fyffe, and Noah Snavely.
\newblock Stereo magnification: Learning view synthesis using multiplane images.
\newblock \emph{arXiv preprint arXiv:1805.09817}, 2018.

\end{thebibliography}
}


\end{document}